\documentclass[conference]{IEEEtran}
\ifCLASSINFOpdf
\else
\fi
\usepackage{graphicx}
\usepackage{setspace}
\usepackage{amsmath}
\usepackage{amssymb} 
\usepackage{bm} 
\usepackage{cite}
\usepackage{float}
\usepackage[usenames,dvipsnames]{pstricks}
\usepackage{epsfig}
\usepackage{pst-grad} 
\usepackage{pst-plot} 
\usepackage{tabularx}
\usepackage[T1]{fontenc}
\newcommand{\lb} {\left}
\newcommand{\rb} {\right}
\newcommand{\nn} {\nonumber}

\allowdisplaybreaks

\begin{document}
\title{Secrecy Outage of Proactive Relay Selection by Eavesdropper}

\author{
    \IEEEauthorblockN{Sarbani Ghose\IEEEauthorrefmark{1}, Chinmoy~Kundu\IEEEauthorrefmark{2}, 
    and Octavia A. Dobre \IEEEauthorrefmark{3}
    }
    \IEEEauthorblockA{\IEEEauthorrefmark{1}Cryptology and Security Research Unit,
Indian Statistical Institute, Kolkata 700108, India }
    \IEEEauthorblockA{\IEEEauthorrefmark{2}School of Electronics, Electrical Engineering and Computer Science, Queen's University Belfast, U.K.}
     \IEEEauthorblockA{\IEEEauthorrefmark{3}Faculty of Engineering and Applied Science, Memorial University, St. John's, NL A1B 3X5, Canada}
    \textrm{\IEEEauthorrefmark{1}sarbani\_v@isical.ac.in}, {\IEEEauthorrefmark{2}c.kundu@qub.ac.uk},
        {\IEEEauthorrefmark{3} odobre@mun.ca}
        }

\maketitle


\begin{abstract}
In this paper, we consider an active eavesdropping scenario in a cooperative system 
consisting of a source, a destination, and an active eavesdropper with multiple 
decode-and-forward relays. Considering an existing assumption in which an eavesdropper is also a part of network, a proactive relay selection by the eavesdropper is proposed. The best relay which maximizes the eavesdropping rate is selected by the eavesdropper. A relay selection scheme is also proposed to improve the secrecy of the system by minimizing the eavesdropping rate. Performances of these schemes are compared with two passive eavesdropping scenarios in which the eavesdropper performs selection and maximal ratio combining on the relayed links. A realistic channel model with independent non-identical links between nodes and 
direct links from the source to both the destination and eavesdropper are assumed. 
Closed-form expressions for the secrecy outage probability (SOP) of these schemes in Rayleigh fading channel are obtained. It is shown that the relay selection by the proactive eavesdropper is most detrimental to the system as not only the SOP increases with the increase in the number of relays, but its diversity also 
remains unchanged.

\end{abstract}

\begin{IEEEkeywords}
Active eavesdropping, cooperative system, Decode-and-forward relay, relay selection, secrecy outage probability.
\end{IEEEkeywords}

\section{Introduction}
\label{sec_intro}
Due to the inherent openness of wireless medium, wireless networks  are vulnerable to 
illegitimate eavesdropping. Over recent years, physical layer security has become 
a promising technique to secure wireless communications by utilizing the physical channel 
properties of both legitimate and illegitimate users \cite{wyner_wiretap, hellman_gaussian_wiretap,
McLaughlin_wireless_info_theo_sec}.



Based on the classical wiretap channel framework, various cooperative transmission techniques 
using relays have been developed against eavesdropping. 
Relays can cooperate mainly in three schemes or a mixture of these schemes: 
i) cooperative beamforming \cite{Petropulu_Poor_Impr_Wire_Phylay_Sec}, 
ii) cooperative jamming \cite{Goel_Negi_Guaranteeing_Secrecy}, 
and iii) relay selection  \cite{krikidis_iet_opport_rel_sel, 
krikidis_twc_Rel_Sel_Jam, Zou_Wang_Shen_optimal_relay_sel, 
Alotaibi_Relay_Selection_MultiDestination, Bao_Relay_Selection_Schemes_Dual_Hop_Security, 
Poor_Security_Enhancement_Cooperative, 
Fan_Karagiannidis_Secure_Multiuser_Communications, 
Kundu_relsel, Qahtani_relsel, kundu_globecom16, fan_Exploiting_Direct_Links}. 
In  cooperative beamforming, all relays perform distributed beamforming towards the
legitimate user to transmit their received signal. In  cooperative jamming, they 
transmit interference or jamming signal to confuse the illegitimate user. These 
techniques require complex signal processing to find beamforming or jamming signal 
vectors at the transmitters or receivers.

Deviating from the former two cases, relay selection schemes select simply the best 
relay (based on different definitions of being the best relay) among several relays 
to forward the received signal towards the legitimate user \cite{krikidis_iet_opport_rel_sel, 
krikidis_twc_Rel_Sel_Jam, Zou_Wang_Shen_optimal_relay_sel, 
Alotaibi_Relay_Selection_MultiDestination, Bao_Relay_Selection_Schemes_Dual_Hop_Security, Poor_Security_Enhancement_Cooperative, 
Fan_Karagiannidis_Secure_Multiuser_Communications, 
Kundu_relsel, Qahtani_relsel, kundu_globecom16, fan_Exploiting_Direct_Links}. 
Based on global instantaneous channel state information (ICSI) or statistical channel 
state information (SCSI) of the links, optimal and suboptimal relay selection schemes are 
developed. It is also shown that some of these relay selection schemes can achieve full 
secrecy diversity. 
A major assumption while achieving global ICSI in these works is that the eavesdropper 
is another active user in the wireless network \cite{McLaughlin_wireless_info_theo_sec, Khandaker_wong_zheng_Truth_Telling_Mechanism}, 
and hence, logically there is no difference between a legitimate or 
illegitimate user capability. It is also to be noted that in all previously mentioned literature, 
relay selection schemes are developed against eavesdropping to safeguard the legitimate communication.     

The literature above assumes that eavesdroppers only listen to a transmission 
intended for the legitimate user. Recently, a new kind of adversary is introduced in secrecy, 
which proactively attacks the system to enhance its eavesdropping performance 
\cite{Amariucai_Wei_Half_duplex_active_eavesdropping, Mukherjee_Swindlehurst_Jamming_games, Xu_Zhang_Proactive_Eavesdropping, 
Zhou_Hjorungnes_Pilot_contamination, 
Zeng_Zhang_ACTIVE_EAVESDROPPING_SPOOFING, Zeng_Zhang_Wireless_Information_Surveillance, 
Tugnait_Detection_of_Active_Eavesdropping, 
Kapetanovi_Rusek_PLS_for_Massive_MIMO, Wu_Schober_Caire_Secure_Massive_MIMO}.  
In \cite{Amariucai_Wei_Half_duplex_active_eavesdropping, Mukherjee_Swindlehurst_Jamming_games, Xu_Zhang_Proactive_Eavesdropping}, 
a jamming signal is transmitted to degrade the received signal quality of the legitimate user when simultaneously eavesdropping. 
While \cite{Amariucai_Wei_Half_duplex_active_eavesdropping, Mukherjee_Swindlehurst_Jamming_games} 
study methods to counter active eavesdropping, \cite{Xu_Zhang_Proactive_Eavesdropping} 
utilizes proactive eavesdropping for surveillance.
In pilot contamination attack \cite{Zhou_Hjorungnes_Pilot_contamination}, the eavesdropper sends 
identical deterministic pilots as the legitimate receiver in the channel training phase of a multi-antenna 
time-division duplexing systems, as opposed to random jamming signal.  The beamformer designed by the 
transmitter on this basis will significantly compromise information to the eavesdropper. In the spoofing relay 
attack proposed in \cite{Zeng_Zhang_ACTIVE_EAVESDROPPING_SPOOFING, Zeng_Zhang_Wireless_Information_Surveillance}, 
the eavesdropper acts as a relay to spoof the source to change its rate in favour of eavesdropping. Detection of such spoofing attack is studied in \cite{Tugnait_Detection_of_Active_Eavesdropping}.  Active eavesdropping 
is also studied in massive multiple-input multiple-output system in \cite{Kapetanovi_Rusek_PLS_for_Massive_MIMO, Wu_Schober_Caire_Secure_Massive_MIMO}.

In the above proactive eavesdropping literature, the eavesdropper jams, contaminates the pilot, or spoofs 
the desired information signal from the source; however, it does not actively select the information source. In a cooperative network where multiple relays are used to convey the message from an information source, 
an eavesdropper can proactively select the best relay to its benefit.  Being a part of the network, an 
eavesdropper either can pretend to be a legitimate destination and thereby compromise the secrecy 
of the system, or can be a legitimate user with a surveillance objective. An eavesdropper proactively 
selecting a relay and its consequences on the secrecy of the system has not been studied so far in the literature.

Motivated by the above discussion, we have considered 
a cooperative network where multiple decode-and-forward (DF) relays can be used by a source 
to convey its message to a destination. An eavesdropper proactively selects the best relay to 
increase its eavesdropping information rate. A realistic scenario is considered where both the 
destination and eavesdropper receive the direct communication from the source; however,  the links between nodes are independent non-identically distributed.

The main contribution of this paper in three fold: 
\begin{enumerate}
  \item We propose a proactive relay selection by the eavesdropper to enhance its eavesdropping capability. 
  Such a scenario is is not considered yet in the literature;
 \item We also study two passive eavesdropping schemes, where the eavesdropper selects the best relayed link in one scheme and maximally combines all the relayed links 
 in the another scheme for comparison with the proposed active eavesdropping. In addition, a passive relay selection scheme based on the
 eavesdropper's channel quality is studied to improve secrecy from the system's point of view for the comparison purpose;
 \item We consider a more general channel model where links between nodes are independent non-identically distributed, as well as with direct 
 links from the source to  the destination and eavesdropper. \footnote{Most of the literature in relay selection to improve security assumes independent 
identical links, without direct links from the source to the destination and eavesdropper 
\cite{krikidis_iet_opport_rel_sel, 
krikidis_twc_Rel_Sel_Jam, Zou_Wang_Shen_optimal_relay_sel, 
Alotaibi_Relay_Selection_MultiDestination, Bao_Relay_Selection_Schemes_Dual_Hop_Security, 
Poor_Security_Enhancement_Cooperative, 
Fan_Karagiannidis_Secure_Multiuser_Communications, 
Qahtani_relsel}. 
Some exceptions can be found in \cite{Kundu_relsel, kundu_globecom16, fan_Exploiting_Direct_Links}, which 
consider either direct links or non-identical links or both.} 
\end{enumerate}


The remainder of the paper is organized as follows. Section \ref{sec_sys_model} 
describes the system model. Section \ref{sec_SOP} evaluates the secrecy
outage probability for different proactive and passive eavesdropping schemes. 
The numerical results are presented in Section \ref{sec_Results}, and conclusions are drawn in 
Section \ref{sec_conclusion}. 

\textit{Notation:} $\mathbb{P}[\cdot]$ is the probability of occurrence 
of an event, $\mathbb{E}_X[\cdot]$ defines the expectation of its argument over the random variable (r.v.) 
$X$, $[x]^+\triangleq \max(0,x)$ and $\max{(\cdot)}$ denotes the maximum 
of its argument, $F_X (\cdot)$ represents the 
cumulative distribution function (CDF) of the r.v. $X$, and 
$f_X (\cdot)$ is the corresponding probability density function (PDF).

\section{System Model}
\label{sec_sys_model}
\begin{figure}
\centering
\includegraphics[width=0.22\textwidth] {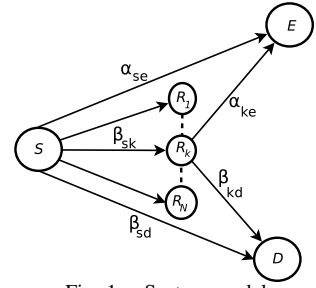}
\vspace*{-0.4cm}
\caption{System model.}
\label{FIG_1}
\vspace{-0.3cm}
\end{figure}

We consider a cooperative system with a source ($S$), a destination ($D$), $N$ DF relays ($R_k$, $k\in \{1,\cdots, N \}$) 
and an eavesdropper ($E$), as shown in Fig. \ref{FIG_1}. There exists direct communication links from $S$ to $D$ and $E$. 
All the links between nodes are modeled as independent non-identically distributed Rayleigh fading channel affected by circularly symmetric independent 
additive white Gaussian noise (AWGN). 
The signal-to-noise ratio (SNR) of the links between $S$-$R_k$, $R_k$-$D$, $S$-$D$, $R_k$-$E$, and $S$-$E$ are denoted as 
$\gamma_{sk}$, $\gamma_{kd}$, $\gamma_{sd}$, $\gamma_{ke}$, and  $\gamma_{se}$. As the links are Rayleigh faded, corresponding SNRs are exponentially distributed with parameters $\beta_{sk}$, $\beta_{kd}$, 
$\beta_{sd}$, $\alpha_{ke}$, and $\alpha_{se}$, respectively, 
where $\mathbb{E}\{ \gamma\}=1/\beta$. \footnote{The subscript denoting a particular path is omitted to show the general case.}  
In the first time slot when $S$ broadcasts its message, it is received by 
$R_k$, $\forall k$, $D$ and $E$. In proactive relay selection by eavesdropper, $E$ selects the best relay which maximizes its information rate 
or equivalently its link SNR and directs it to retransmit in the second time slot.

The achievable secrecy rate of the system is defined as \cite{wyner_wiretap, hellman_gaussian_wiretap}, 
\begin{align}
\label{eq_1}
C_s\triangleq{\frac{1}{2}\lb[\log_2\lb(\frac{1+\gamma_{M}}{1+\gamma_{E}}\rb)\rb]}^+,
\end{align} 
where $\gamma_{M}$ and $\gamma_{E}$ are the SNRs at $D$ and $E$, respectively, after maximal-ratio-combining (MRC) the direct and relayed transmission in the second time slot. 
The multiplier $1/2$ is for the required two time slots to complete the communication process.
For the DF relay system, the rate of a dual-hop link is limited by the
minimum of the individual hop rates. This is equivalent to the minimum of the individual hop SNRs. 
The secrecy rate of a single DF relay, i.e., only for the $k$th relay, following \cite{Zou_Wang_Shen_optimal_relay_sel, Kundu_relsel} is
\begin{align}
\label{eq_5}
C_s^k  = {\frac{1}{2}\lb[\log_2\lb(\frac{1+\gamma_{kD}+\gamma_{sd}}{1+\gamma_{ke}+\gamma_{se}}\rb)\rb]}^+,
\end{align}
where $\gamma_{kD}=\min\lb(\gamma_{sk},\gamma_{kd}\rb)$. The distribution of $\gamma_{kD}$ is also exponential with parameter $\beta_{kD}=\beta_{sk}+\beta_{kd}$.
%
%
%

\section{Secrecy Outage Probability}
\label{sec_SOP}
The secrecy outage probability (SOP) is defined as the probability that achievable secrecy capacity 
is below a certain threshold secrecy rate 
\begin{align}
\label{eq_2}
P_o(R_s)&=\mathbb{P}\lb[C_s<R_s\rb]  
=\mathbb{P}\lb[\frac{1+\gamma_M}{1+\gamma_E} \leq \rho \rb], 
\end{align}
where $R_s$ is the threshold secrecy rate and $\rho=2^{2R_s}$.

\subsection{Proactive Relay Selection by $E$ (MAX-E)}
\label{subsec_max_eve} 
In this scheme, $E$ selects the $k$th relay for which the $R_k$-$E$ link SNR is maximum. 
Once the relay is selected by $E$, $R_k$ decodes the signal received from $S$ and broadcasts it. 
The SOP of the system can be derived from \eqref{eq_2} by finding the probability of the selection of the best relay 
and then finding the SOP if the particular relay is selected. By summing up all, 
SOP is obtained following the law of total probability as  
\begin{align}
\label{eq_8}
&P_o(R_s)
=\sum_{k=1}^N\mathbb{P}\lb[\gamma_{ke}>\gamma_{ke}^{-}\rb] 
\mathbb{P}\lb[\frac{1+\gamma_{sd}+\gamma_{kD}}{1+\gamma_{se}+\gamma_{ke}}<\rho \rb], 
\end{align}
where
$\gamma_{ke}^{-}=\max_{\substack{i=1,\dots,N\\i\ne k}}\{ \gamma_{ie}\}$ denotes the maximum SNR among the rest of the channels received by $E$ except the $k$th one.

 

The PDF of $\gamma_{ke}^{-}$ can be obtained from the CDF of 
$\gamma_e=\max_{\substack{k=1,\dots,N}}\{ \gamma_{ke}\}$ following \cite{Kundu_relsel}  as
\begin{align}
\label{eq_4}
F_{\gamma_e}(x)=1+\sum\limits_{m=1}^N(-1)^m \mathbb {\sum}_m e^{-x\alpha_m^{\prime}},
\end{align}
where 
\begin{align}
\label{eq_4_A}
\mathbb {\sum}_m=\sum_{i_1=1}^{N-(m-1)}\sum_{i_2=i_1+1}^{N-(m-2)}\cdots\sum_{i_{m-1}=i_{m-2}+1}^{N-1} \sum_{i_m=i_{m-1}+1}^N, 
\end{align}
and $\alpha_m^{\prime}=\sum_{l=1}^m\alpha_{i_l e}$.
The PDF of the r.v. $\gamma_{ke}^{-}$ can be expressed in the easily realizable summation form as 
\begin{align}
\label{eq_9}
 f_{\gamma_{ke}^-}(x)=-\sum\limits_{m=1}^{N-1}\lb(-1\rb)^{m}\mathbb {\sum}_m^{\prime} \alpha_m^{\prime}e^{-x\alpha_m^{\prime}},
\end{align} 
where 
\begin{align}
\label{eq_10}
\mathbb {\sum}_m^{\prime}=\sum_{\substack{i_1=1\\i_1\ne k}}^{N-(m-1)}
\sum_{\substack{i_2=i_1+1\\i_2\ne k}}^{N-(m-2)}\cdots \sum_{\substack{i_{m-1}=i_{m-2}+1\\i_{m-1}\ne k}}^{N-1} 
\sum_{\substack{i_m=i_{m-1}+1\\i_m\ne k}}^N.
\end{align}
Then, we can evaluate \eqref{eq_8} as follows 
\begin{align}
\label{eq_14}
&P_o(R_s)\nn\\
&=\sum_{k=1}^N\mathbb{P}\lb[\gamma_{ke}>\gamma_{ke}^{-}\rb]\mathbb{P} \lb[\gamma_{ke}> \frac{\gamma_{sd}+\gamma_{kD}-\rho\gamma_{se}-\lb(\rho-1\rb)}
{\rho}\rb]\nn\\
&=\sum_{k=1}^N\lb[I_1+I_2+I_3\rb].
\end{align}
Assuming $X=\gamma_{kD}+\gamma_{sd}$, $Z=\gamma_{se}$ and 
$\lambda
=\frac{x-z\rho-(\rho-1)}{\rho}$, finally $I_1$, $I_2$ and $I_3$ are expressed in the integral forms in \eqref{eq_MAX_E_def1}, \eqref{eq_MAX_E_def2}, and \eqref{eq_MAX_E_def3}, 
respectively. For the distribution of the r.v. $X$, which is the sum of two independent non-identically exponentially distributed r.v.s, we follow \cite{sum_expo_mohamed_akkouchi},
where 
$ B_1 = \frac{\beta_{kD} \beta_{sd}}{\beta_{kD}-\beta_{sd}}$ and
$B_2 = \frac{\beta_{kD} \beta_{sd}}{\beta_{sd}-\beta_{kD}}$.

\subsection{Relay Selection by the System to Minimize Eavesdropping Rate (MIN-E) }
\label{subsec_min_eve} 
The eavesdropper in this section is assumed to be passive. The system selects the relay to 
enhance its secrecy by choosing the lowest quality link to $E$ only. 
This means that the relay having minimum instantaneous SNR among all the links tapped by $E$ is selected. 
Following the law of total probability, the SOP can be obtained as
\begin{align}
\label{eq_34}
&P_o(R_s)
=\sum_{k=1}^N\mathbb{P}\lb[\gamma_{ke}<\gamma_{ke}^{-}\rb] 
\mathbb{P}\lb[\frac{1+\gamma_{sd}+\gamma_{kD}}{1+\gamma_{se}+\gamma_{ke}}<\rho \rb] \nn\\
&=\sum_{k=1}^N\mathbb{P}\lb[\gamma_{ke}<\gamma_{ke}^{-}\rb] 
\mathbb{P}\lb[\gamma_{ke}>\frac{\gamma_{sd}+\gamma_{kD}-\rho\gamma_{se}-(\rho-1)}{\rho}\rb], 
\end{align}
where
$\gamma_{ke}^{-}=\min_{\substack{i=1,\dots,N\\i\ne k}}\{ \gamma_{ie}\}$.

The PDF of the r.v. $\gamma_{ke}^-$ can be easily obtained from the definition of CDF and 
differentiating it as
\begin{align}
\label{eq_pdfmin}
f_{\gamma_{ke}^-}(y)=\alpha \exp \lb(-\alpha y \rb), 
\end{align}
where $\alpha= \sum_{\substack{i=1\\i\ne k}}^N \alpha_{ie}$.
Then (\ref{eq_34}), can be evaluated as
\begin{align}
\label{eq_subse1} 
P_o (R_s)&=\sum_{k=1}^N
\mathbb{P}\lb[\frac{\gamma_{kD}+\gamma_{sd}-\rho\gamma_{se}-(\rho-1)}{\rho} \le{\gamma_{ke}}\le{\gamma_{ke}}^-\rb]\nn\\
&=\sum_{k=1}^N \lb[ I_4+I_5\rb].
\end{align}
For the derivation above, we assume $X=\gamma_{kD}+\gamma_{sd}$ and $\lambda=\frac{x -\rho z-(\rho-1)}{\rho}$.
$I_4$ and $I_5$ can be expressed in the integral form in \eqref{eq_pdfmin2_def} and \eqref{eq_pdfmin3_def}, respectively. Their solutions can be stated in \eqref{eq_pdfmin2} and \eqref{eq_pdfmin3}, 
where $B_1=\frac{\beta_{kD}\beta_{sd}}{\beta_{kD}-\beta_{sd}}$ and 
$B_2=\frac{\beta_{kD}\beta_{sd}}{\beta_{sd}-\beta_{kD}}$ are due to the PDF of $X$. 

\subsection{Best Relayed Link Selection at $E$ while MRC at D (MAX-MRC)}
\label{subsec_mrc_max}
In this section $E$ is assumed to be a passive listener. In the first time slot, $E$ simply chooses the best relayed link among all, and performs MRC with the direct transmission. No relay selection is involved. $D$ performs MRC with all the relayed and direct links received by it in 
the first and second time slots.
From \eqref{eq_2}, we obtain the SOP of the system as 
\begin{align}
\label{eq_28}
&P_o(R_s)
=\mathbb{P}\lb[\frac{1+\gamma_{sd}+\sum_{k=1}^N \gamma_{kD}}
{1+\gamma_{se}+ \gamma_e}<\rho\rb] \nn \\
&=1-\mathbb{P}\lb[\gamma_e\leq \frac{\sum_{k=0}^N \gamma_{kD}-\rho\gamma_{se}-\lb(\rho-1\rb)}{\rho}\rb]\nn\\
&=1-\int_{0}^\infty\int_{\rho z+\rho-1}^\infty F _Y 
\lb(\lambda\rb) 
f_X(x)dx f_Z(z) dz. 
\end{align}
For the derivation above, we assume 
$\gamma_e= \max_{\forall k } \gamma_{ke}$, $\gamma_{0D}=\gamma_{sd}$, $Y=\gamma_e$, $X=\sum_{k=0}^N \gamma_{kD}$, 
$Z=\gamma_{se}$, and $\lambda=\frac{x-z\rho-(\rho-1)}{\rho}$.
For the PDF of $X$, we follow the PDF of the sum of $N+1$ independent non-identically distributed exponential r.v.s from \cite{sum_expo_mohamed_akkouchi} as,
\begin{align}
\label{eq_25}
f(t)=\sum_{i=1}^N\frac{\prod_{j=1}^N \lambda_j}
{\prod_{\substack{j=1 \\
 j\ne i}}^N\lb(\lambda_j-\lambda_i\rb)} e^{-t\lambda_i},
\end{align} 
where $\lambda_i$, $\forall i$, are the parameters of the exponential distribution.
We substitute the CDF of $\gamma_e$ and PDF of $X$ from \eqref{eq_4} 
and \eqref{eq_25}, respectively, in \eqref{eq_28}
to obtain the SOP expression in \eqref{eq_sop_mrc_max}.

\subsection{MRC of All Relayed Links with the Direct Links at E and D (MRC-MRC)}
\label{subsec_mrc_mrc}
This section discusses the SOP when $E$ and $D$ both combine signals received in 
the first and second time slots using MRC. The SNRs at 
$D$ and $E$  are
$\gamma_{M} 
= \gamma_{sd} + \sum_{k=1}^N  \gamma_{kD}$ and
$\gamma_{E}=
\gamma_{se} + \sum_{k=1}^N \gamma_{ke}$, respectively. 
The SOP of the MRC-MRC technique can be obtained from \eqref{eq_2} as 
\begin{align}
\label{eq_27} 
&P_o(R_s)
=\int_0^\infty F_{\gamma_M}\lb(\rho x+\rho-1\rb) f_{\gamma_E}(x)dx.
\end{align}
For the distribution of $\gamma_M$ and $\gamma_E$, we use the PDF of the sum of $N+1$ independent non-identically distributed r.v.s 
\cite{sum_expo_mohamed_akkouchi}. 
Substituting the CDF and PDF, SOP can be evaluated as
\begin{align}
\label{eq_27A} 
P_o(R_s)
&=\int_0^\infty \sum_{i=1}^{N+1} \sum_{p=1}^{N+1} 
\frac{\prod_{\substack{j=1 \\ 
j\ne i}}^{N+1}\beta_{jD}
\prod_{\substack{q=1 
}}^{N+1}\alpha_{qe}}
{\prod_{\substack{j=1 \\ 
j\ne i}}^{N+1}\lb(\beta_{jD}-\beta_{iD}\rb)
\prod_{\substack{q=1, \\ 
q\ne p}}^{N+1}\lb(\alpha_{qe}-\alpha_{pe}\rb)} \nn \\
&\times\lb(1-\exp\lb[-\beta_{iD}\lb(\rho-1+\rho x\rb)\rb]\rb)e^{-\alpha_{pe}x} dx, 
\end{align}
whose solution is shown in \eqref{eq_sop_mrc_mrc}.

\section{Numerical Results}
\label{sec_Results}

\begin{figure}
\centering
\includegraphics[width=0.5\textwidth] {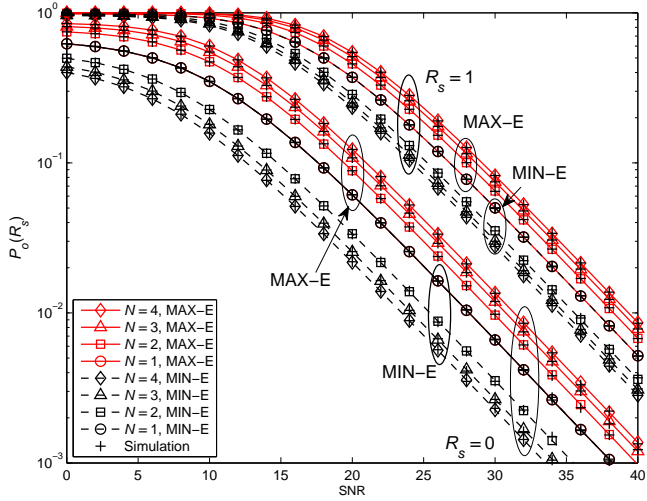}
\vspace*{-0.4cm}
\caption{SOP versus SNR of MAX-E and MIN-E for different $N$ and $R_s$ values, with 
$1/\alpha_{ke}=3$ dB, $\forall k$, 
$1/\beta_{sd} = 3$ dB, 
$1/\alpha_{se} = 0$ dB, 
and $1/\beta_{sk} = 1/\beta_{kd}$.
}
\label{FIG_2}
\vspace{-0.3cm}
\end{figure}

\begin{figure}
\centering
\includegraphics[width=0.48\textwidth] {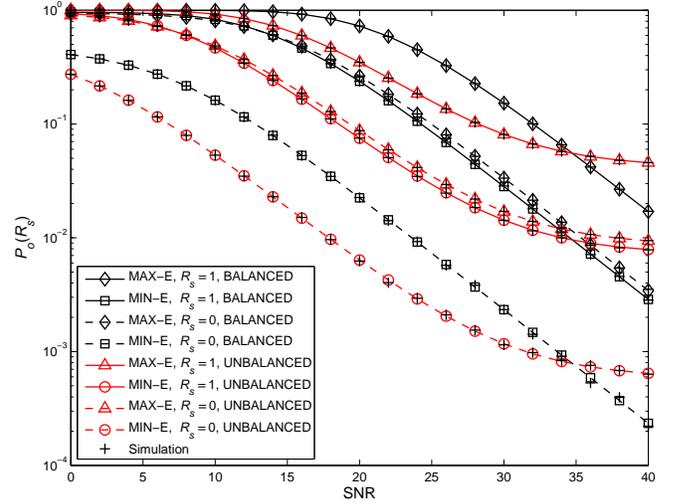}
\vspace*{-0.4cm}
\caption{SOP versus SNR of MAX-E and MIN-E in  balanced ($1/\beta_{sk} = 1/\beta_{kd}$)
and unbalanced case ($1/\beta_{sk} = 30$ dB)
 with $1/\alpha_{ke} = \{0, 3, 6, 9\}$ dB, $k=1, \cdots, 4$,  
$1/\beta_{sd} = 3$ dB, 
$1/\alpha_{se} = 0$ dB. 
}
\label{FIG_3}
\vspace{-0.3cm}
\end{figure}

\begin{figure}
\centering
\includegraphics[width=0.48\textwidth] {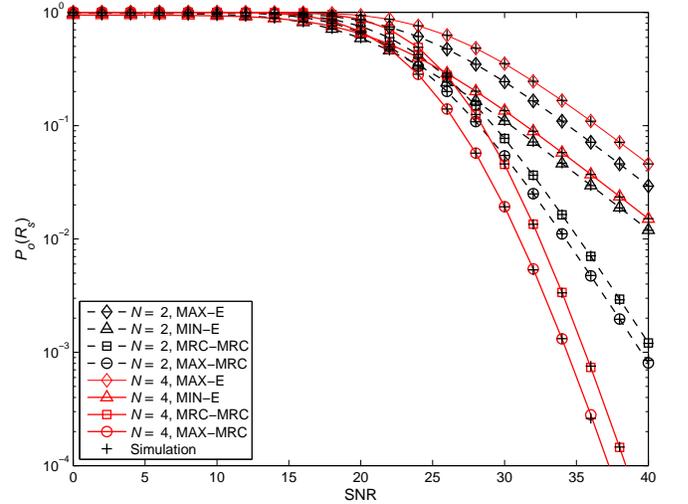}
\vspace*{-0.4cm}
\caption{SOP versus SNR of all schemes for 
$N=2, 4$ with $R_s = 1$ bpcu, $1/\beta_{sd} = 3$ dB, 
$1/\alpha_{se} = -3$ dB, and having all other channel parameters non-identical. 
}
\label{FIG_4}
\vspace{-0.3cm}
\end{figure}

In this section, numerical results are plotted along with simulation results to verify the analytical findings.
It is assumed that the AWGN affects all receiving nodes equally.
Unless otherwise specified, $P_o(R_s)$ is plotted against SNR with SNR divided equally between $S-R_k$ and $R_k-D$ , $\forall k$. 
The direct link average SNRs are considered as $1/\beta_{sd} = 3$ dB and $1/\alpha_{se} = 0$ dB. 
The required threshold $R_s$ is assumed either $0$ or $1$ bits per channel use (bpcu). 
Secrecy outage probability with $R_s=0$ provides the intercept probability by $E$, as 
an intercept event occurs when the secrecy capacity becomes negative 
\cite{Zou_Wang_Shen_optimal_relay_sel}.
These particular parameter values are only for the illustration purpose. 
Other values provide the same trend in the figure; results are not included here due to space limitations.

Fig. \ref{FIG_2} compares the proposed relay selection schemes, MAX-E and MIN-E, by increasing $N$ from 1 to 4. 
Eavesdropping link average SNRs are considered equal with $1/\alpha_{ke}=3$ dB, $ \forall k$. 
It can be observed that not only $P_o(R_s)$ is worse for MAX-E, but it also increases with the increase in $N$, while it decreases 
for MIN-E. MAX-E is exactly the requirement from the $E$'s perspective; however, it is detrimental to the system which tries to avoid eavesdropping. 
Moreover, it can be seen that performances do not change equally with the increase in $N$. The change is more when $N$ is increased from $1$ to $2$; thereafter, the changes diminish with subsequent increase in $N$. 
In addition, it is noticed that though $N$ increases, the slopes of the curves do not change.
Furthermore, it is also observed that SOP increases with the increase in $R_s$.
However, $N$ affects less at higher $R_s$ as curves for $R_s=1$ are more closely spaced than $R_s=0$.

Fig. \ref{FIG_3} compares MAX-E and MIN-E for $N=4$ when the eavesdropping link average SNRs are all different, i.e., 
$1/\alpha_{ke}=0$, $3$, $6$, and $9$ dB, respectively, for $k=1, \cdots 4$. Two cases are depicted: balanced and unbalanced. In the balanced case, 
$1/\beta_{sk}=1/\beta_{kd}$, $\forall k$, whereas, in the unbalanced case $1/\beta_{sk}= 30$ dB, $\forall k$,  $1/\beta_{kd}$, $\forall k$, 
increases as SNR in the $x$-axis. It can be observed that in the balanced case, the performances improve with SNR; on the contrary, in the unbalanced case, 
the performances saturate depending on the SNR of the $S-R_k$  links. In the DF relay systems, the dual hop SNR is constrained by the minimum of the 
individual hop SNRs, and hence, the observation. The observation would have been similar 
if $1/\beta_{kd}$ had been fixed to a particular SNR as well.

Fig. \ref{FIG_4} compares the performances of all schemes when each of the $S-R_k-D$ 
branch effective SNRs, 
as well as all the eavesdropping link average SNRs are different for $N=2$ and $N=4$, respectively. 
When $N=2$, $1/\alpha_{ke}= 6, 9$ dB for $k=1, 2$, whereas when $N=4$, $1/\alpha_{ke}= 0, 3, 6, 9$ dB, for  $k=1, \cdots, 4$.
$1/\alpha_{se}=-3$ dB is considered to make it different from other eavesdropping links.
A balanced case, i.e., $1/\beta_{sk}=1/\beta_{kd}$ is considered. 
When $N=2$, $1/\beta_{sk}$ is assumed to be 20\% and 30\% for $k=1, 2$, respectively, of the SNR representing $x$-axis. When $N=4$, $1/\beta_{sk}$ is assumed to be 5, 10, 15, and 20\% for $k=1, \cdots, 4$, respectively, to make the total $S-R_k-D$, $\forall k$, link SNRs 100\%. 

It is observed that for a given $N$, the proactive relay selection by $E$, MAX-E, is the worst for the system, although the best for $E$.
Though, both MAX-E and MAX-MRC select the maximum eavesdropping channel quality for them, the 
proactive eavesdropping is better than 
the passive eavesdropping from $E$'s perspective.  The best scheme for the system secrecy is MAX-MRC as $D$ performs MRC combining.
From $E$'s perspective, MRC-MRC is better than MAX-MRC as $E$ is able to perform MRC on the relayed signals. 

Although the systems with $N=2$ and $N=4$ are not comparable as parameters are different, it can be noticed that the secrecy diversity order-the slope of the curves at high SNR-remains the same for both MAX-E and MIN-E.  $P_o(R_s)$ remains parallel 
as $N$ increases, as seen in Fig. \ref{FIG_2}. On the contrary, both passive eavesdropping scheme MRC-MRC and MAX-MRC benefit from the increase in $N$, as the
secrecy diversity order improves to aid the system secrecy. This concludes that proactive eavesdropping on the basis of the relay 
selection is the best for $E$. Moreover, the relay selection scheme, MIN-E,  to aid the system secrecy based on the minimum among the 
eavesdropping link quality, performs worse than both MAX-MRC and MRC-MRC.

Simulation results exactly match the theoretical ones, which confirms the validity of 
the analytical findings.

\section{Conclusion}
\label{sec_conclusion}
Relay selection is thus far considered to improve the secrecy of the system. Being part of a wireless 
network, an eavesdropper might pretend to be a legitimate destination, and actively 
select a relay to maximize its  eavesdropping capability. We study such an active eavesdropping scenario, 
where the eavesdropper selects the best relay to maximize the eavesdropping rate. Three other passive eavesdropping scenarios are also considered focusing on the eavesdropping link and eavesdropping capability for the comparison purpose. Closed-form expressions for the secrecy outage probability are obtained considering a more realistic scenario of independent no-identically distributed channels and the
direct links between the source and both destination and eavesdropper. It is concluded that the active eavesdropping is the worst for the system secrecy, and the secrecy outage probability decreases with the increase in the number of relays, while the secrecy diversity remains unchanged.

\section*{Acknowledgment}
\label{sec_acknowledgment}
This work was supported in part by the Royal Society-SERB Newton International Fellowship under Grant NF151345, 
and by the Natural Science and Engineering Council of Canada (NSERC), through its Discovery program.

\begin{table*}
\begin{align}
\label{eq_MAX_E_def1} 
I_1 &=\int_0^\infty \int_{\rho z+\rho-1}^\infty \int_0^\lambda \int_\lambda^\infty  
f_{\gamma_{ke}}(t)f_{\gamma_{ke}^{-}}(y)f_X(x)f_{\gamma_{se}}(z)dtdydxdz  \\
\label{eq_MAX_E_1} 
&=-\sum\limits_{m=1}^{N-1}(-1)^m \mathbb {\sum}_m^{\prime} \rho \alpha_{se}
\lb(\frac{B_1 e^{-\beta_{sd}\lb(\rho-1\rb)}}
{\lb(\alpha_{ke}+\rho\beta_{sd}\rb)\lb(\alpha_{se}+\rho\beta_{sd}\rb)}
+\frac{B_2 e^{-\beta_{kD}\lb(\rho-1\rb)}}
{\lb(\alpha_{ke}+\rho\beta_{kD}\rb)\lb(\alpha_{se}+\rho\beta_{kD}\rb)} \rb. \nn \\
&\lb.-\frac{B_1 e^{-\beta_{sd}\lb(\rho-1\rb)}}
{\lb(\alpha_{ke}+\alpha_m^{\prime}+\rho\beta_{sd}\rb)\lb(\alpha_{se}+\rho\beta_{sd}\rb)}
-\frac{B_2 e^{-\beta_{kD}\lb(\rho-1\rb)}}
{\lb(\alpha_{ke}+\alpha_m^{\prime}+\rho\beta_{kD}\rb)\lb(\alpha_{se}+\rho\beta_{kD}\rb)}\rb) .\\
\label{eq_MAX_E_def2} 
I_2&=\int_0^\infty \int_{\rho z+\rho-1}^\infty \int_\lambda^\infty \int_y^\infty
f_{\gamma_{ke}}(t)f_{\gamma_{ke}^{-}}(y)f_X(x)f_{\gamma_{se}}(z)dtdydxdz  \\
\label{eq_MAX_E_2} 
&= -\sum\limits_{m=1}^{N-1}(-1)^m \mathbb {\sum}_m^{\prime} \frac{\rho\alpha_m^{\prime}\alpha_{se}}{\alpha_{ke}+\alpha_m^{\prime}} 
\lb(\frac{B_1 e^{-\beta_{sd}(\rho-1)}}
{\lb(\alpha_{ke}+\alpha_m^{\prime}+\rho\beta_{sd}\rb)\lb(\alpha_{se}+\rho \beta_{sd}\rb)} 
+\frac{B_2 e^{-\beta_{kD}(\rho-1)}}
{\lb(\alpha_{ke}+\alpha_m^{\prime}+\rho\beta_{kD}\rb)\lb(\alpha_{se}+\rho \beta_{kD}\rb)} \rb). \\
\label{eq_MAX_E_def3} 
I_3&=\int_0^\infty \int_0^{\rho z+\rho-1} \int_0^\infty \int_y^\infty
f_{\gamma_{ke}}(t)f_{\gamma_{ke}^{-}}(y)f_X(x)f_{\gamma_{se}}(z)dtdydxdz \\
\label{eq_MAX_E_3} 
&=-\sum\limits_{m=1}^{N-1}(-1)^m \mathbb {\sum}_m^{\prime} \frac{\alpha_m^{\prime}\alpha_{se}}{\alpha_{ke}+\alpha_m^{\prime}}
\lb[\frac{B_1}{\alpha_{se}\beta_{sd}}+\frac{B_2}{\alpha_{se}\beta_{kD}}
-\frac{B_1e^{-\beta_{sd}(\rho-1)}}{\beta_{sd}\lb(\alpha_{se}+\rho\beta_{sd}\rb)}
-\frac{B_2e^{-\beta_{kD}(\rho-1)}}{\beta_{kD}\lb(\alpha_{se}+\rho\beta_{kD}\rb)}\rb] .
\end{align}
\hrule 
\begin{align}
\label{eq_pdfmin2_def}
I_4 &=\int_{0}^\infty \int_{\rho z+\rho -1}^\infty \int_\lambda^\infty \int_\lambda^y 
f_{\gamma_{ke}}(t)f_{\gamma_{ke}^{-}}(y)f_{X}(x)f_{\gamma_{sd}}(z) dt dy dx dz \\
\label{eq_pdfmin2}
&=\frac{\alpha_{se}\alpha_{ke}}{\alpha+\alpha_{ke}}\lb[\frac{B_1\exp\lb(-\beta_{sd}(\rho-1)\rb)}
{\lb(\rho\beta_{sd}+\alpha_{se}\rb)\lb((\alpha_{ke}+\alpha)/\rho+\beta_{sd}\rb)} 
+\frac{B_2\exp\lb(-\beta_{kd}(\rho-1)\rb)}{\lb(\rho\beta_{kd}+\alpha_{se}\rb)\lb((\alpha_{ke}+\alpha)/\rho+\beta_{kd}\rb)}  \rb]. \\
\label{eq_pdfmin3_def}
I_5 &= \int_{0}^\infty \int_{0}^{\rho z+\rho -1}\int_0^\infty \int_0^y 
f_{\gamma_{ke}}(t)f_{\gamma_{ke}^{-}}(y)f_{X}(x)f_{\gamma_{sd}}(z) dt dy dx dz \\
\label{eq_pdfmin3}
&=\frac{\alpha_{ke}}{\alpha+\alpha_{ke}}\lb[\frac{B_1}{\beta_{sd}}+\frac{B_2}{\beta_{kD}}
-\frac{B_1\alpha_{se}\exp\lb(-\beta_{sd}(\rho-1)\rb)}{\lb(\rho\beta_{sd}+\alpha_{se}\rb)\beta_{sd}} 
-\frac{B_2\alpha_{se}\exp\lb(-\beta_{kD}(\rho-1)\rb)}{\lb(\rho\beta_{kD}+\alpha_{se}\rb)\beta_{kD}}  \rb].
\end{align}
\hrule
\begin{align}
\label{eq_sop_mrc_max}
P_o(R_s)
&=1-\alpha_{se} \lb(\sum_{i=1}^{N+1}
\frac{\prod_{j=1}^{N+1}\beta_{jD}}
{\prod_{\substack{j=1 \\ 
j\ne i}}^{N+1}\lb(\beta_{jD}-\beta_{iD}\rb)}\rb) 
\lb[\frac{e^{-\beta_{iD}(\rho-1)}}{\beta_{iD}\lb(\alpha_{se}+\rho\beta_{iD}\rb)}
-\sum\limits_{m=1}^{N}(-1)^m 
\mathbb {\sum}_m 
\frac{\rho e^{-\beta_{iD}(\rho-1)}}{\lb(\alpha_m^{\prime}+\rho\beta_{iD}\rb)\lb(\alpha_{se}+\rho\beta_{iD}\rb)} 
\rb].
\end{align}
\hrule
\begin{align}
\label{eq_sop_mrc_mrc}
P_o(R_s)
 &=\sum_{i=1}^{N+1} \sum_{p=1}^{N+1}
\frac{\prod_{\substack{j=1 \\ 
j\ne i}}^{N+1}\beta_{jD}
\prod_{\substack{q=1 
}}^{N+1}\alpha_{qE}}
{\prod_{\substack{j=1 \\ 
j\ne i}}^{N+1}\lb(\beta_{jD}-\beta_{iD}\rb)
\prod_{\substack{q=1, \\ 
q\ne p}}^{N+1}\lb(\alpha_{qe}-\alpha_{pe}\rb)} 
\lb(\frac{1}{\alpha_{pe}}-\frac{e^{-(\rho-1)\beta_{iD}}}{\alpha_{pe}+\rho\beta_{iD}}\rb).
\end{align}
\hrule
\end{table*}

  \bibliographystyle{IEEEtran}
  \bibliography{IEEEabrv,MYALL_REFERENCE}

\end{document}